\documentclass[sigconf]{acmart}

\usepackage{enumitem}
\usepackage{color,soul} %highlight text
\usepackage[disable]{todonotes}% enable to show responses
\usepackage{xcolor} 
\usepackage{subcaption}

\makeatletter
\if@todonotes@disabled
\newcommand{\hlfix}[2]{#1}
\else
\newcommand{\hlfix}[2]{\texthl{#1}\todo{#2}}
\fi
\makeatother

\AtBeginDocument{%
  \providecommand\BibTeX{{%
    \normalfont B\kern-0.5em{\scshape i\kern-0.25em b}\kern-0.8em\TeX}}}

\copyrightyear{2022} 
\acmYear{2022} 
\setcopyright{acmcopyright}\acmConference[CHI '22]{CHI Conference on Human Factors in Computing Systems}{April 29-May 5, 2022}{New Orleans, LA, USA}
\acmBooktitle{CHI Conference on Human Factors in Computing Systems (CHI '22), April 29-May 5, 2022, New Orleans, LA, USA}
\acmPrice{15.00}
\acmDOI{10.1145/3491102.3517528}
\acmISBN{978-1-4503-9157-3/22/04}

\begin{document}

\title[Enhancing Community Resilience among Older Adults through Crisis Informatics Tools]{Investigating Older Adults’ Attitudes towards Crisis Informatics Tools: Opportunities for Enhancing Community Resilience during Disasters} 

\author{Nurul M Suhaimi} 
\orcid{0000-0002-4318-5805}
\affiliation{%
  \institution{Northeastern University}
  %\streetaddress{360 Huntington Ave}
  \city{Boston}
  \state{MA}
  \postcode{}
  \country{USA}
}
\affiliation{%
  \institution{Universiti Malaysia Pahang}
  %\streetaddress{360 Huntington Ave}
  \city{}
  \state{Pahang}
  \postcode{}
  \country{Malaysia}
}
\email{suhaimi.n@northeastern.edu}

\author{Yixuan Zhang} 
\orcid{0000-0002-7412-4669}
\affiliation{%
  \institution{Georgia Institute of Technology}
  %\streetaddress{801 Atlanta Dr}
  \city{Atlanta}
  \state{GA}
  \postcode{}
  \country{USA}
}
\email{yixuan@gatech.edu}

\author{Mary Joseph} 
%\orcid{0000-0000-0000-0000}
\affiliation{%
  \institution{Northeastern University}
  %\streetaddress{360 Huntington Ave}
  \city{Boston}
  \state{MA}
  \postcode{}
  \country{USA}
}
\email{joseph.mary@northeastern.edu}

\author{Miso Kim} 
%\orcid{0000-0000-0000-0000}
\affiliation{%
  \institution{Northeastern University}
  %\streetaddress{360 Huntington Ave}
  \city{Boston}
  \state{MA}
  \postcode{}
  \country{USA}
}
\email{m.kim@northeastern.edu}

\author{Andrea G Parker}
\orcid{0000-0002-2362-7717}
\affiliation{%
  \institution{Georgia Institute of Technology}
  %\streetaddress{801 Atlanta Dr}
  \city{Atlanta}
  \state{GA}
  \postcode{}
  \country{USA}
}
\email{andrea@cc.gatech.edu}

\author{Jacqueline Griffin} 
\orcid{0000-0001-7729-1748}
\affiliation{%
  \institution{Northeastern University}
  %\streetaddress{360 Huntington Ave}
  \city{Boston}
  \state{MA}
  \postcode{}
  \country{USA}
}
\email{ja.griffin@northeastern.edu}

\renewcommand{\shortauthors}{Suhaimi et al.}

\begin{abstract}
    The world population is projected to rapidly age over the next 30 years. Given the increasing digital technology adoption amongst older adults, researchers have investigated how technology can support aging populations. However, little work has examined how technology can support older adults during crises, despite increasingly common natural disasters, public health emergencies, and other crisis scenarios in which older adults are especially vulnerable. Addressing this gap, we conducted focus groups with older adults residing in coastal locations to examine to what extent they felt technology could support them during emergencies. Our findings characterize participants’ desire for tools that enhance community resilience-local knowledge, preparedness, community relationships, and communication, that help communities withstand disasters. Further, older adults' crisis technology preferences were linked to their sense of control, social relationships, and digital readiness. We discuss how a focus on community resilience can yield crisis technologies that more effectively support older adults.
\end{abstract}


\maketitle

\section{Introduction}
By 2050, the number of adults aged 65 and above in the world will be 1.5 billion, double the numbers in 2017~\cite{world2019united}. In the United States, this shift will increase the median age by ten years~\cite{pew2014aging}. \hlfix{The increase in the older adult population will indirectly increase the number of at-risk populations during emergencies, since older adults are more likely than others in a community to have multiple chronic conditions, limitations in daily activities, and disabilities that impede their ability to communicate about, prepare for, and respond to a natural disaster}{Rewritten to address R2's minor comment \#1}~\cite{shih2018improving}. For instance, older adults made up 75 percent of the fatalities in Hurricane Katrina~\cite{brunkard2008hurricane}, \hlfix{and 73 percent of the deaths from the October 2017 Wildfires in Northern California}{Rewritten to address R2's minor comment \#2}~\cite{liu2015systematic,badger2018catastrophic}. Additionally, in the recent COVID-19 pandemic, mortality rates among adults aged 60 and above were the highest among age groups~\cite{bonanad2020effect}, not only due to results of illness~\cite{brooke2020older}, but also resulting from isolation and mobility challenges~\cite{wu2020social,benksim2020vulnerability}.

While there has been significant progress toward understanding the disaster risk profile among older adults~\cite{campbell2019disaster, tuohy2016older}, there are significant research gaps that must be addressed in order to continue advancing support for older adults in times of crisis. For example, few studies consider the abilities and experiences of older adults living independently in community settings (i.e., community-dwelling older adults)~\cite{campbell2019disaster}. Additionally, most research does not highlight the active role older adults play in making independent decisions pertaining to their safety~\cite{campbell2019disaster, zhang2020understanding}. Instead, much of the existing literature focuses on a limited view that considers older adults as passive victims in emergencies~\cite{campbell2019disaster}. 

Similar to how current literature disregards the active participation of older adults in crisis response, to the best of our knowledge, there is minimal research that focuses on the use of digital technology by this population in disasters (see \autoref{sec:literature_intersect}). While many researchers found that older adults are increasingly using information and communication technologies (ICTs)~\cite{bloomberg2020, pruchno2019technology, pew2021internet}, and that ICTs play a significant role in disaster scenarios~\cite{haddow2013disaster, apuke2018social, gambura2018two, gray2018supporting}, there remains a lack of understanding about how well these tools can serve older adults, which can introduce intervention-generated inequalities~\cite{veinot2018good}. 

Prior work that has explored the use of technology during crises (i.e., crisis informatics tools) among older adults has examined features supporting their sense of control, dignity, and safety~\cite{zhang2020understanding}. Yet, these features have focused on older adults as \emph{individuals} rather than promoting their engagement with \emph{communities}. Simultaneously, crisis informatics tools tend to focus on addressing individuals' needs rather than strengthening communities, even though the key features of crisis informatics tools are to enhance connections and communications during crises~\cite{wei2006staying, zhao2019understanding}. \hlfix{Instead, crisis informatics tools for older adults must address both their unique needs and facilitate opportunities to maximize the full potential of older adults' active roles in the safety of themselves and their communities.}{Revised to remove patronizing language}

To address this research gap, we conducted \hlfix{focus groups with older adults residing in coastal locations to examine to what extent and how they felt technology could support them during emergencies. The focus groups consisted of three parts: 1) a semi-structured focus group discussion, 2) a storyboard activity and discussion, and 3) a survey assessment}{Rewritten to address meta-review \#1}  Specifically, our work is guided by the following research questions (RQs):
\begin{enumerate}%[label=RQ \arabic*:]
\item[RQ1:] {What are older adults' attitudes towards emergencies (e.g., perceived barriers, decision-making strategies) and towards community support during emergencies?}
\item[RQ2:] {What are older adults' preferences for and attitudes towards crisis informatics tools that facilitate community suppor in emergencies?}
\end{enumerate}

To answer our research questions, we conducted a study involving adults aged 65 years and older to investigate their attitudes and preferences towards crisis informatics tools which provide support during emergencies. Our findings characterize participants' desire for tools \hlfix{that reflect the key elements of}{Rewritten to address R3's comment \#1} community resilience--local knowledge, preparedness, community relationships, and communication~\cite{patel2017we}. Here, we define \textit{community resilience}, in alignment with Andrew, as ``communities and individuals harnessing local resources and expertise to help themselves in an emergency, in a way that complements the response of the emergency services''~\cite{andrew2012building}. In addition, prior work has identified how older adults' abilities to react in crises is dependent on their sense of control~\cite{zhang2020understanding} and social connectedness~\cite{wei2006staying}. Therefore, in this study, we assess how these features correspond to our participants' preferences for crisis informatics tools. We discuss how a focus on community resilience can yield crisis technologies that more effectively support older adults. 

\hlfix{This work contributes to HCI research in the following ways. First, our work adds to the literature in crisis informatics by providing new insights into the perceptions of older adults pertaining to the usefulness and perceived values of crisis informatics tools that support community resilience for older adults, focusing on four elements--local knowledge, preparedness, community relationships, and communication. Second, our work offers a counternarrative to the societal portrayal of older adults as primary receivers of care and support by providing evidence of the critical roles they play in society during crisis. Finally, we discuss the opportunities and challenges in developing crisis informatics tools for older adults that include the cost of submitting personal information on these tools.}{Added to address R1's comment \#3}

%------------------------------------------------------------

\section{Related Work} 
\label{sec:literature}

In the following sections, we describe the definition of community resilience, particularly as it pertains to crises, disasters, and emergencies. We focus on research that is situated at the intersection of community resilience and crisis informatics. We also discuss the evolving partnership between older adults and technology, as well as prior research that addresses this partnership for emergencies.

\subsection{Crisis Informatics \& Community Resilience} \label{sec:literature_resilience}

Crisis informatics is an interdisciplinary research area that examines the interconnectedness of people, organizations, information, and technology during crises~\cite{hagar2010crisis}. Research in crisis informatics emphasizes the networked digital technology used in times of crisis and how technology enables interaction across places and among people. A central topic of crisis informatics research is crisis and risk communication, which heavily relies on sociotechnical systems. Specifically, there has been a growing interest in understanding the interactions between people, organizations, institutions, and a range of technologies in entangled heterogeneous arrangements in which ``social'' and ``technical'' components of systems cannot be isolated in practice~\cite{lamb2000social, zhang2022shifting}. 

Research on crisis informatics has examined a wide range of topics, spanning from emergency management and operations, to community resiliency and self-reliance~\cite{palen2007crisis}. We discuss the body of work on resilience, as it is relates to our research focus. From a social-ecological perspective, resilience can be defined as the ``capacity of a system to absorb disturbance and reorganize to remain essentially the same function, structure, identity, and feedbacks''~\cite{walker2004resilience, folke2010resilience}. Building upon the concept of absorbing and reorganizing and the effects of climate change, the concept of ``community resilience'' against a crisis or disaster is highly sought after by emergency response professionals, government officials, and academics~\cite{patel2017we}. While the concept of community resilience is almost invariably viewed as positive, the exact definition of community resilience is still debated in scientific literature, policies, and practices~\cite{patel2017we}. Nonetheless, a systematic review conducted by Patel et al. found that the general consensus regarding the elements of community resilience, as it applies to disasters, includes local knowledge, preparedness, community relationships, and communication~\cite{patel2017we}. Following these elements, in this study, we define \textit{community resilience} as ``communities and individuals harnessing local resources and expertise to help themselves in an emergency, in a way that complements the response of the emergency services''~\cite{andrew2012building}. 

\hlfix{Driven by the growing proportion of adults aged 65 and above, researchers have called for research on community resilience to pay more attention to older adults, which would be conducive to the realization of successful aging~\cite{cohen2016community,wiles2012resilience,rand2011building, zhang2021community}. For example, older adults with a strong sense of community can contribute their experience, resources, and relationship-building capacity to support others during emergencies~\cite{shih2018improving,zhang2021community}. In turn, older adults can both generate and mobilize social capital at the local level during disasters~\cite{shih2018improving, zhang2021community,richards2021hugs}. By designing tools that allow older adults to stay engaged and exert their own choices~\cite{tuohy2016older,zhang2020understanding}, we can continue to support older adults emotional health~\cite{light2016s} and resiliency~\cite{gibson2013expanding,madsen2019enhancing,shih2018improving}, as well as expanding their social connectedness~\cite{richards2021hugs}.}{Added to address R3's comment \#10}  

In this work, we focus on how community resilience, as it pertains to its four core elements (local knowledge, preparedness, community relationships, and communication), can be enhanced with crisis informatics tools that reflects upon older adults' preferences. This addresses a key research gap, where minimal crisis informatics research has focused on community resilience among traditionally vulnerable populations, such as older adults. Further,  Soden and Palen  have stated how the narrow temporal framing in drawing attention to the role of technologies can perpetuate structural inequities in identifying who is affected and who is able to recover~\cite{soden2018informating}.

\subsection{Intersection of Older Adults, Technology, and Crisis} \label{sec:literature_intersect}

To our knowledge, the intersection of older adults, technology, and crisis in a single study is limited. From our point of view, two factors are responsible for this limitation--(1) the slow growth of technology use among older adults, and (2) the perception that older adults' technological needs during disasters are similar to that of other age groups. Here, we provide the reasoning behind these factors and discuss how these factors are not only limited by their temporality but also compromised by cohort change. 

Since the advent of information and communication technology, particularly in the introduction of smartphones~\cite{Anderson2017tech} and the propagation of social media~\cite{pew2018mobilefact}, the partnership between older adults and technology has been evolving~\cite{pruchno2019technology}. For example, a recent Pew Research Center study that found only 7\% of older adults do not have access to technology in 2021, compared to 47\% in 2000~\cite{pew2021internet}. Looking forward, by the time the current age group of 40-60 years old enters the ``older adult'' cohort, they will bring their higher levels of technology use with them, and we can expect greater use of technology, specifically with smartphones usage, among the population 65 years and above.

\hlfix{Correspondingly, a growing number of studies have sought to develop mobile applications that enhance the quality of life of the older adults population~\cite{klimova2017smartphone, kim2022healthy, helbostad2017mobile} and support their capacity to age well~\cite{durick2013dispelling, light2015ageing, giaccardi2018resourceful}. These studies have described how older adults were willing to and able to learn and use tools that they recognized as useful and which support their sense of control and independence~\cite{light2015ageing, giaccardi2018resourceful}. Yet, many of these studies continue to focus on older adults day-to-day lives, rather than crisis events such as natural disasters, fire emergencies, and pandemics that may affect their lives significantly.}{Added to address R3's comment \#6} Thus, the projected increase in technology uptake among older adults must be used to identify and develop mobile applications that support the unique needs of older adults during disasters. 

For example, the significant progress in understanding older adults' needs and vulnerabilities during disasters~\cite{tuohy2016older, oriol1999psychosocial, ngo2001disasters}, need to be translated into tools that support older adults' in times of crises. To the best of our knowledge, limited studies have considered coupling these tools with technology that augment support and enable autonomy among older adults during disasters. Notable works include a study by Gibson et al. who discussed the potential for technological solutions to combat the disproportionate vulnerability of older adults in all phases of emergency management~\cite{gibson2013expanding}. Ashida et al. \hlfix{developed a personal disaster preparedness program by identifying the characteristics of emergency support networks and providers sought by older adults}{Rewritten to address R2's minor comment \#3}~\cite{ashida2017personal}. A study by Zhang et al. has explored the aspects of crisis apps (i.e., mobile applications for crisis) that support older adults' sense of control, safety, and dignity during crises~\cite{zhang2020understanding}. \hlfix{A recent study by Richards et al. has identified multiple technology modalities used by older adults to stay resilient during COVID-19 pandemic, specifically those used for connecting with others~\cite{richards2021hugs}. The same study has also called for designing technology that fosters smart relationships--relationships that are in familiar forms of touch such as hugs, relationships that facilitate new connections, and relationships that are creative, playful, and spontaneous, virtually~\cite{richards2021hugs}.}{Added to address meta-review \#8} However, little work examines how crisis informatics tools support community resilience among older adults pertaining to the four elements described above. Thus, our work contributes to the body of literature at the intersection of older adults, technology, and crisis, by exploring how crisis informatics tools can be designed to enhance community resilience among older adults.

%------------------------------------------------------------

\section{Method} \label{sec:method}
To answer our research questions, we conducted \hlfix{focus groups with 18 older adults consisting of three parts: 1) a semi-structured focus group discussion, 2) a storyboard activity and discussion, and 3) a survey assessment.}{Rewritten to to address meta-review \#1 to improve the clarity about methods} Below we detail our recruitment and study procedures\hlfix{\footnote{The study protocol was approved by our university's Institutional Review Board.}}{Written as footnote to address meta-review \#2}.

\subsection{Recruitment \& Study Participants}
Participants were recruited in coastal towns in the Northeastern United States through local organizations that provide community outreach to older adults. Eligible participants were required to be 65 years or older, capable of completing all study components (e.g., \hlfix{group discussions}{Changed from interviews for clarity/address meta-review \#1} and survey) in English, had past experience with at least one emergency (i.e., hurricane, tornado, snowstorm, etc.), and used technology for communication daily (i.e., smartphones, tablets, laptops, etc.). Our sample (n = 18) included 14 women and 4 men. Their ages ranged between 65 and 86 (median = 76). \autoref{tab:demographic} presents an overview of our participants' demographic information.

\begin{table*}[ht!]
  \caption{Participant demographic information}
  \label{tab:demographic}
\tabcolsep20pt
  \begin{tabular}{cccccc}%{cccccc}
    \toprule
        \shortstack{Participant\\ ID} & \shortstack{Age\\ Range} & Gender & \shortstack{Highest Education\\ Level} & \shortstack{Marital\\ Status} & \shortstack{Employment \\ Status}\\ 
    \midrule
    P01 & 65-69 & Female & Bachelor Degree      & Single  & Part-time\\
    P02 & 70-74 & Female & Bachelor Degree      & Single  & Part-time\\
    P03 & 75-79 & Female & Bachelor Degree      & Single  & Unemployed\\
    P04 & 80-84 & Female & Bachelor Degree      & Single  & Retired\\
    P05 & 70-74 & Female & Associate's Degree   & Single  & Retired\\
    P06 & 70-74 & Male   & Vocational Training  & Married & Retired\\
    P07 & 80-84 & Male   & Bachelor Degree      & Single  & Self-employed\\
    P08 & 70-74 & Female & Vocational Training  & Married & Retired \\
    P09 & 75-79 & Male   & Graduate Degree      & Single  & Retired\\
    P10 & 70-74 & Female & Bachelor Degree      & Single  & Retired\\
    P11 & 65-69 & Female & Graduate Degree      & Single  & Retired\\
    P12 & 75-79 & Female & Graduate Degree      & Single  & Full-time\\
    P13 & 85+   & Male   & Graduate Degree      & Married & Self-employed\\
    P14 & 75-79 & Female & Bachelor Degree      & Single  & Retired\\
    P15 & 75-79 & Female & High School          & Single  & Retired\\
    P16 & 75-79 & Female & Bachelor Degree      & Single  & Retired\\
    P17 & 65-69 & Female & Bachelor Degree      & Married & Self-employed\\
    P18 & 75-79 & Female & Graduate Degree      & Single  & Retired\\
  \bottomrule
\end{tabular}
\end{table*}

\subsection{Study Procedure}
\begin{figure*}[ht!]
\centering
\includegraphics[width=1.0\textwidth]{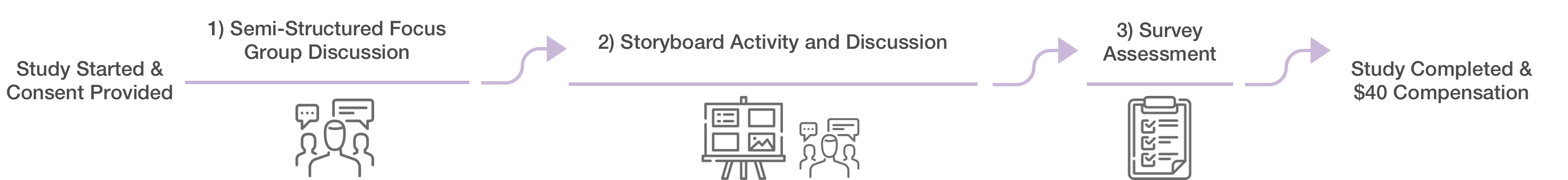}
\caption{An overview of our study flow: Part 1 consisted of a semi-structured focus group discussion (see \autoref{sec:method_focus_group}). Part 2 involved a storyboard activity (see \autoref{sec:method_participatory_design}). Part 3 had a survey assessment (see \autoref{sec:method_survey}). The study was completed after the survey assessment and participants were compensated for their time.}
\label{fig:study_flow}
\end{figure*}

We conducted this study in February 2020, \hlfix{approximately one month before COVID-19 was declared a pandemic by the World Health Organization (WHO)~\cite{who2020covid} and a national emergency was declared in the United States~\cite{whitehouse2020covid}. Correspondingly, safety measures such as social distancing and mask usage were not yet enacted at the time of our study.}{Added to address meta-review \#2 and R2's comment \#1} Upon arrival, participants were briefly introduced to the purpose of the study by the researchers and signed a consent form allowing for all discussions to be audio recorded. \hlfix{Participants were then divided into four groups of 4-5 people.}{Added to address address meta-review \#1 and R3's comment \#3} Each focus group followed the same protocol and contained three parts. The first part focused on discussions about participants' experiences with disasters and their general perceptions of local emergency services. The second part involved a storyboard activity and discussion. In the last part, each participant independently completed a survey. \autoref{fig:study_flow} \hlfix{briefly illustrates our study flow.}{Added a new flow diagram to address meta-review \#1 and R3's comment \#2}  

\subsubsection{\hlfix{Part 1: Semi-Structured Focus Group Discussion}{Separated for clarity/ to address meta-review \#1}}
\label{sec:method_focus_group}
The discussion in each focus group commenced with a short introduction and a brief explanation of the session by the moderator. Each participant was given some time to ask any questions about the session and to introduce themselves. A semi-structured guide was used for the focus group discussion. The first question was designed to be an opening, easy-to-answer question to encourage all the participants to talk and feel comfortable. Specifically, participants were asked about their current neighborhood and the amount of time they have lived there. To address our RQ1, participants were asked about the unique challenges of living in their neighborhood, specifically in a coastal region, which is vulnerable to natural disasters~\cite{kusenbach2010disaster, kafle2012measuring}. They were also asked about their awareness of emergency services available in their communities and their perceptions about the reliability of local resources for providing assistance during emergencies. The participants were always given enough time to discuss each question thoroughly until there were no more responses. Once all questions were asked and all opinions were expressed, the moderator thanked the participants and moved onto the next portion of the session. 

\subsubsection{\hlfix{Part 2: Storyboard Activity and Discussion}{Separated for clarity and address meta-review \#1}}
\label{sec:method_participatory_design}
\begin{figure*}[htp!]
\centering
\includegraphics[width=0.8\textwidth]{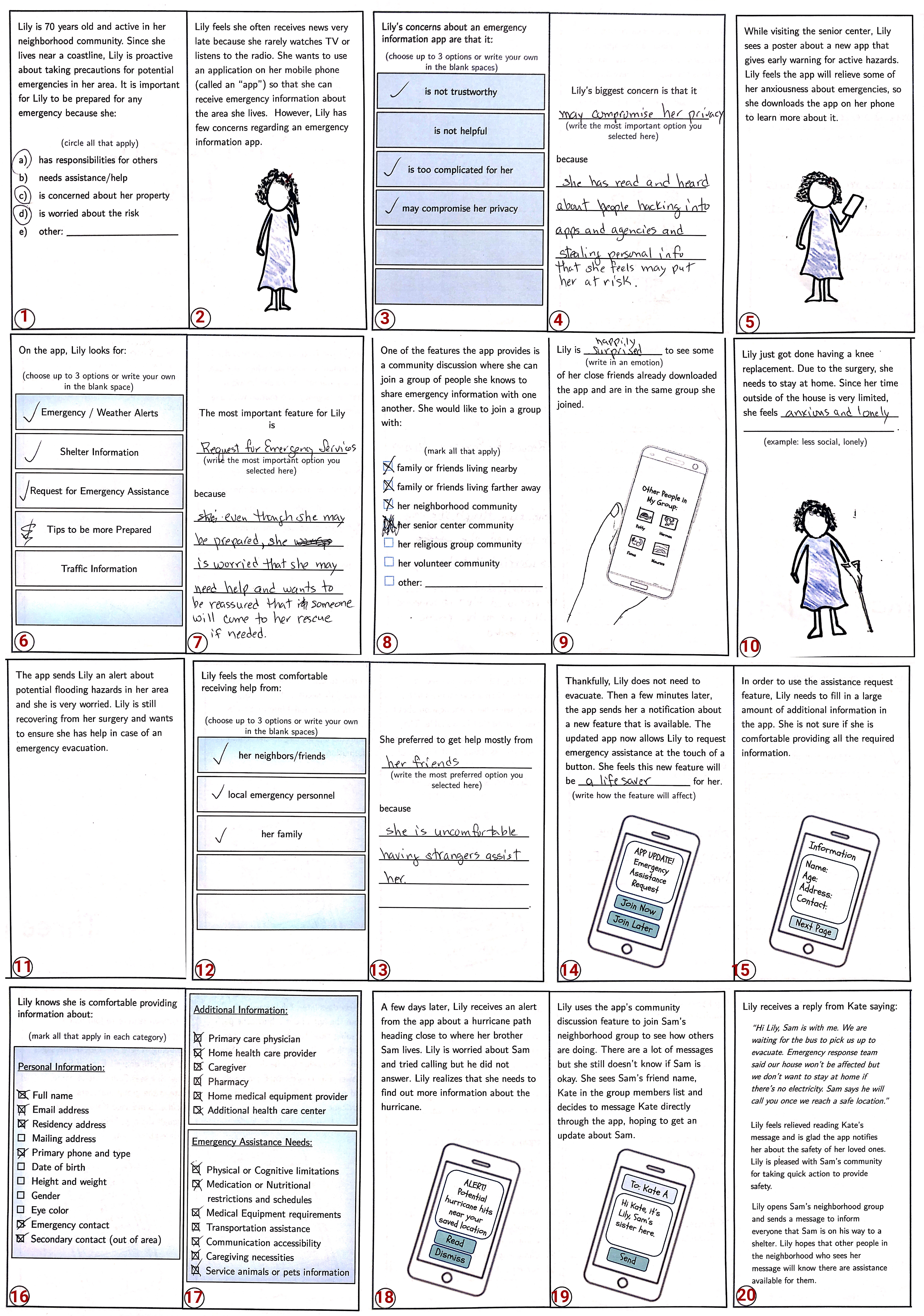}
\caption{An example of a completed storyboard with the order of the storyboard scenes labeled from (1) to (20).} 
\label{fig:storyboard}
\end{figure*}

\hlfix{In the second part of the focus group, participants were asked to complete a storyboard (see \autoref{fig:storyboard}) to explore their attitudes towards and preferences for crisis app features that enhance community support (to answer RQ2). Similar to Stowell et al. who utilized a modular storyboard approach to evaluate mobile health applications~\cite{stowell2020investigating}, our modular storyboard allowed us to stimulate participants' evaluation of and ideation around mobile applications used for crisis events. Modular storyboards are a hybrid form of storyboard, incorporating elements from two traditional types of storyboards--blank canvas~\cite{halko2010personality} and completed storyboards~\cite{berry2019supporting}. This hybrid form of storyboard integrate the benefits of providing participants with blank canvas storyboards (e.g., allowing them the freedom to convey their own design visions) and the benefits of showing participants completed storyboards (e.g., providing them with an idea of what is possible in the design space). With modular storyboards, participants are given partially-completed storyboards and then guided through a process of sharing their reactions to the storyboard, modifying and completing the storyboard in a way that conveys their vision for how a future technology should be designed and used, and why. By using a modular storyboard approach, our participants were able to customize parts of the storyboards individually, while still having all participants react to the same overarching storyline (i.e., narrative plot that describes a sequence of main events)~\cite{stowell2020investigating}. Taking this approach thus enabled participants to express their personal needs, values, and unique experiences as they relate to crisis apps. At the same time, the common storyline facilitated group discussions and comparisons of participants' boards.}{Rewritten to address meta-review \#1 and R3's comment \#3} 

The storyboards depicted a fictional character’s interactions with a crisis app before, during, and after a disaster. \hlfix{Specifically, we integrated crisis app features that support older adults' sense of control, dignity, and safety, based on the findings by Zhang et al.~\cite{zhang2020understanding} in the storyboard. A storyline that aligns with the features depicted three phases of the character's crisis app journey: introduction to the app (Scenes 1 to 5 in \autoref{fig:storyboard}), using the features of the app (Scenes 6 to 10), and seeking support from the crisis app (Scenes 11 to 20). At each phase of the character's journey, the storyboard options allowed the participants to create a crisis app that was authentic to their own needs and experiences. Participants were instructed to customize the storyboards by selecting from multiple-choice options or filling-in-the-blank (see \autoref{fig:storyboard}). For example, in Scene 8 of the storyboard, participants were asked to choose the group of people they would like to connect with through the crisis app. Options such as ``family and friends living nearby'', ``family and friends living farther away'', and ``neighborhood community'' were provided, along with a blank space for options that were not listed. In other scenes, participants were asked to narrate the character's emotions, attitudes and preferences to help them further convey the rationale behind how they would desire to interact with a crisis app, and why (e.g., see Scenes 4, 7, 9, 13, 14). For some of the scenes in the storyboard, participants were asked to select their most preferred option from the options listed (see Scenes 6 and 7 in \autoref{fig:storyboard}).}{Rewritten to address meta-review \#1 and R3's comment \#3} \hlfix{Through the storyboard, participants were also prompted to identify their willingness to provide information which is often required during emergency response~\cite{newhanover2021} (see Scenes 16 and 17).}{Rewritten to address R2's minor comment \#4}

At the end of the storyboard activity, one participant was randomly chosen to present their storyboard to the other participants in their group, who then discussed similarities and differences with their own storyboards. The group discussions enabled participants to agree with and challenge perspectives and attitudes of one another. At the end of the storyboard discussion, the moderator thanked the participants and moved to the next portion of the session.

The average total time taken for Part 1 and 2 was 110 minutes and ranging between 100 and 120 minutes.

\subsubsection{Part 3: Survey Assessment} \label{sec:method_survey}

\hlfix{Upon completion of Part 1 and 2 of the focus group,}{Replaced ``Following the interview'' and address meta-review \#1} each participant individually completed a survey consisting of four sections: 1) demographic information, 2) sense of control, 3) social relationships, and 4) digital readiness\hlfix{\footnote{All survey questions are available in the supplementary materials.}}{Added to address meta-review \#1, R2's comment \#1, and R3's comment \#4}. \hlfix{Each of these attributes were measured to capture how they relate with participants' choices and preferences for tools that facilitate community support. To identify a set of survey measures, we reviewed the literature for widely used, validated scales that contained strong psychometric properties~\cite{lachman1998sense,teo2013social,horrigan2016digital}. The resulting scales were chosen due to their high reliability in measuring the attributes of focus in our study.}{Added to address meta-review \#1 and R3's comment \#4}

\hlfix{We assessed sense of control since it often relates to an individual's ability to make decisions in times when stress is mounting~\cite{tuohy2016older}. Moreover, prior work has found that the act of engaging with available support resources, a demonstration of sense of control, enhances older adults' resilience during disasters~\cite{madsen2019enhancing,shih2018improving}. Yet, these studies missed to understand how sense of control relates to older adults' preferences for digital tools that support them during crises.}{Revised to address meta-review \#1 and R3's comment \#4} We used a widely employed questionnaire by \hlfix{Lachmann and Weaver}{Added to address R2's minor comment \#6}~\cite{lachman1998sense} to measure participants' sense of control. Participants responded to a series of questions with a 7-point Likert scale (1  - Strongly Disagree, 7 - Strongly Agree) on personal mastery and perceived constraints. Higher scores on personal mastery reflect a greater sense of control and independence (i.e., personal control over one's life circumstances)~\cite{robinson2017perceived}, while higher scores on perceived constraints reflect a greater feeling that one is controlled by external factors (i.e., individuals' self-estimated constraints that prevent completion of tasks or successful goal attainment)~\cite{robinson2017perceived}. Thus, a person with high personal mastery and low perceived constraints is associated with a high sense of control~\cite{lachman1998sense}. 

\hlfix{Social relationships were assessed to gauge the level of connections our participants had with their family and friends, since personal connections are often found to be the primary resource that older adults seek assistance from during disasters~\cite{ashida2017personal}. Further, social relationships and connectedness are found to be closely tied with community resilience~\cite{jewett2021social,cagney2016social}.}{Revised to address meta-review \#1 and R3's comment \#4} The level of connections were measured using social contact questions from the MIDUS survey~\cite{teo2013social}, a national longitudinal cohort study on social relationships. Participants were asked to rate the frequency of being in contact (e.g., visits, phone calls, letters, or emails) with their 1) family members that are not in their household, and 2) friends, on a scale of 1 (Never or hardly ever) to 8 (Several times a day). 

Given our focus on exploring future opportunities for the design of crisis applications, familiarity with technology was critical to ensure that our participants had the background knowledge and experience to engage in dialogue, critique and design ideation around crisis apps. Distinct from prior work~\cite{zhang2020understanding}, our study focus on older adults who used technology on a daily basis. Correspondingly, we include survey questions to assess participants' digital comfort and readiness using a 4-item instrument from a questionnaire used by Pew Research Center~\cite{horrigan2016digital}.

After the completion of the survey, participants were thanked by the researchers and received \$40 cash as compensation for their time.

\subsection{Analysis}\label{sec:methods_analysis}
We utilized the methodological approach of triangulation ~\cite{kobayashi2019international}, combining qualitative and quantitative data to gain a deeper understanding of how participants' \hlfix{responses during group discussions and preferences regarding crisis apps tools}{Replaced ``interview data'' and to address meta-review \#1 and R2's comment \#1} complement their survey results.

\hlfix{For the survey data, we utilized descriptive statistics (average and standard error) to characterize our participants' sense of control, social relationships, and digital readiness. Results from the survey are presented in \autoref{sec:survey_results}. Group discussions were audio-recorded and transcribed verbatim for analysis purposes.}{Rewritten for clarity and to address meta-review \#1 and R2's comment \#1} We conducted an inductive analysis of the transcripts~\cite{thomas2006general} using the NViVo 12 Pro software~\cite{nvivo}. In the coding process, we first employed attribute coding and structural coding, dividing the transcript data roughly into two main categories: needs during emergencies and attitudes towards tools-supported crisis apps. \hlfix{Next, within each category, we read the transcripts line-by-line and created codes to label concepts in the data. We then clustered low-level codes to form higher-level themes, which include the elements of community resilience.}{Revised to address R3's comment \#1} Results from the inductive analysis are presented in \autoref{sec:findings}.

\subsection{Reflexivity \& Positionality}\label{sec:methods_reflexivity}
\hlfix{The authors of this paper include graduate students and professors from different racial and cultural groups. While none of the authors are older adults (65 years and above), all of us have family members and close friends who are in this cohort, and have had experienced emergencies and crises. The first and second author grew up in cities that were frequently hit with natural disasters. As a result of roots and familial ties to these high-risk regions, we feel a strong affinity to the challenges faced by our communities. During times of crises, the first and second author's aging parents and grandparents have utilized mobile apps for informing us on their safety. And yet, they have also faced challenges in using these tools in risk mitigation, and thus, failed in addressing their needs. Our positions offer us a unique perspective when doing research on the ways in which mobile technologies become both enabling and constraining in assisting older adults in times of crises. In addition, we had the privilege to work alongside our team members who include researchers with prior experiences conducting studies at the intersection of older adults and HCI. Leveraging our experiences in these fields, we designed our focus group study, starting with semi-structured, easy-to-answer questions to give voice to those whose views are rarely heard, as well as engaging older adults to co-create technologies that meet their needs and requirements (i.e., storyboard activities). In short, given how communication technologies have been and will continue to play an important role in our day-to-day lives, and since there are rarely mobile applications built with considerations for older adults' needs in times of crises, our research has focused on how technological innovation can be sensitively and effectively created with and for older adults.}{Added to address R3's comment \#5}

%------------------------------------------------------------

\section{Survey Results}\label{sec:survey_results}
\hlfix{To provide background of our participants, we present an analysis from the survey regarding sense of control, social relationships, and digital readiness.}{Separated from Findings section in previous version and to address meta-review \#1, R1’s comment \#5, and R2's comment \#2}

\begin{figure*}[ht!]
\begin{minipage}[b]{0.33\textwidth}
\includegraphics[height=1.5in]{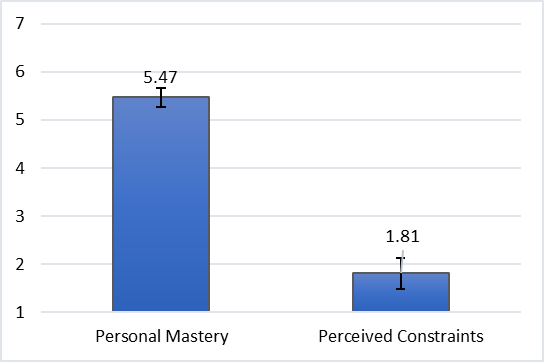}
\subcaption{Sense of control ratings}\label{fig:ratings_control}
\end{minipage}
\hfill
\begin{minipage}[b]{0.33\textwidth}
\includegraphics[height=1.5in]{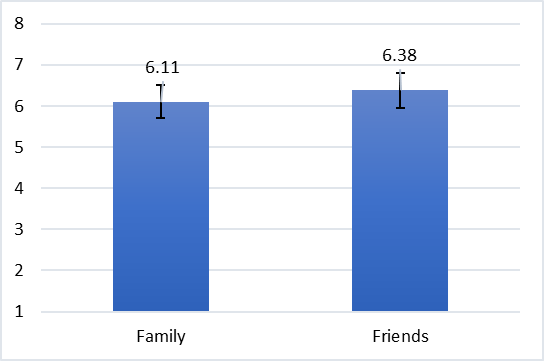}
\subcaption{Social relationships ratings}\label{fig:ratings_contact}
\end{minipage}
\hfill
\begin{minipage}[b]{0.33\textwidth}
\includegraphics[height=1.5in]{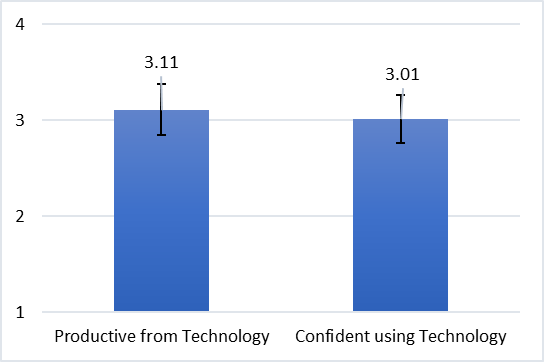}
\subcaption{Digital readiness ratings}\label{fig:ratings_digital}
\end{minipage}%
\caption{Average and standard error ratings by participants on the sense of control~\cite{lachman1998sense}, social relationships~\cite{teo2013social}, and digital readiness ~\cite{horrigan2016digital}. Sense of control is rated from 1 (``strongly disagree'') to 7 (``strongly agree''), social relationships are rated from 1 (``never or hardly ever'') to 8 (``several times a day''), and digital readiness is rated from 1 (``not well at all'') to 4 (``very well'').} \label{fig:ratings}
\end{figure*}

\autoref{fig:ratings_control} shows the average and standard error ratings on personal mastery and perceived constraints for the participants. Overall, our participants reported greater personal mastery (score$\geq$5) and lower perceived constraints (score$\leq$3), demonstrating a high sense of control and independence (see \autoref{sec:method_survey} and list of questions in Appendix). As for social relationships, on average, participants communicated with their family and friends at least once a week (score$\geq$5), as shown in \autoref{fig:ratings_contact}. As we targeted inclusion of older adults with daily technology use, our participants also are classified as ``digitally ready'' (score$\geq$3), where they are comfortable and confident using technology independently.

In the next section, we present how these characteristics link to our participants' disaster preparedness and their preferences towards crisis informatics tools in enhancing community resilience.

%------------------------------------------------------------

\section{Findings}\label{sec:findings}

\hlfix{In this section, we present our findings from the group discussions. Specifically, we examine how older adults' preferences for crisis informatics tools are consistent with the key elements of community resilience: local knowledge, preparedness, community relationships, and communication.}{Rewritten to address R3's comment \#7}

\subsection{Addressing Situated Vulnerability and Preparedness}
As described in \autoref{sec:literature}, community resilience involves four elements---local knowledge, preparedness, community relationships, and communication ~\cite{patel2017we}. \hlfix{In this section, we describe how local knowledge and preparedness are linked to participants' situated vulnerability and preparedness.}{Highlighted to address R3's comment \#7} We define \textit{situated vulnerability} as vulnerability related to spatial and temporal aspects of residing in a particular location.

\subsubsection{Situated Vulnerability}
Vulnerability in older adults has often been described as being related to physical and mental abilities to take actions and make decisions pertaining to one's safety ~\cite{tuohy2016older}. However, our participants described how their vulnerability was not attributed to their physical and mental challenges of being an older adult but instead was attributed to residing in a coastal location with significant seasonal tourism.

Our participants discussed how their geographical situatedness informed their risk perceptions and preparedness. As P08 stated, when describing the challenges of evacuation during crises, \textit{``I think if you have to leave [the area] it would be extremely difficult.''}. Our participants reside in a coastal area in the Northeastern United States which is connected to the mainland via bridges. The unique geographical features of the region limit the ability for emergency evacuation. This limitation was described by P10, \textit{``And the thing is, it's a bridge. If it's a tsunami or terrible winds, we're not going to get over that bridge.''}. 
Further, this limitation varies with the seasons. During summer, the increased tourist population exacerbates the difficulties for evacuation. P06 described the situation from his personal experience:
\begin{quotation}
\textbf{P06}: \textit{``If [the wind is] 70 miles an hour or greater during a blizzard or a hurricane, the bridges are shut down...the evacuation route means nothing because you won't be able to evacuate. Think about what happened when Hurricane Edouard, and Labor Day weekend of 1996... this hurricane had formed, and it was coming straight up and was going to be hitting the [area]. Everyone tried to leave the [area] over Labor Day weekend...and you had tourists, everyone, there was a backup ...  all the way to the bridges.''}
\end{quotation}

While there is a population increase in the summer, in winter, many residents, referred to as \textit{``snow birds''}, leave the region for warmer climates.  Participants who live by themselves described the situation as \textit{``peaceful''}, but \textit{``isolated''}, resulting in a \textit{``lack of immediate support''}. In fact, one participant (P04) described how she was unable to ask for help during a power outage that lasted for several days since her closest neighbors were not in the area and she had limited mobility to travel farther than her house.

The situated vulnerability, as described by P06 and P04, while causing unique challenges, has also influenced participants to actively prepare for crisis events. P03 described:
\begin{quotation}
\textbf{P03}: \textit{``...because we cannot leave this island. As much as we don't want to admit, this is an island, and we're separated from the mainland by those two bridges. So I think our preparedness...we have to not think about evacuating, leaving here, but what we're going to do...how we're going to survive here. So I think our situation ... is different than other parts of [the state].''}
\end{quotation}

P03's description exemplifies the awareness of the situated vulnerability faced by our participants. This awareness has induced the desire to make preparations to shelter-in-place in emergencies (i.e., \textit{``not think about evacuating''}). Next, we describe the actions taken by our participants in their emergency preparations.

\subsubsection{Situated Preparedness}
The effects of a disaster, whether short-term or long-term, can be mitigated if a community understands its existing vulnerabilities~\cite{patel2017we}. These vulnerabilities, if addressed prior to the disaster, can induce the building of resilience within the community~\cite{cohen2016community}. Here, we define \textit{situated preparedness} as the actions taken to prepare for disasters which are influenced by situated vulnerabilities~\cite{baker2014role}.

Extending upon the discussion of situated vulnerability, our participants described various examples of situated preparedness, including having home generators and battery-operated radios. For P06, commonly occurring power  have been mitigated through the use of an \textit{``at-home''} generator. On the other hand, 9 out of 18 participants (P01, P03, P07, P08, P12, P13, P15, P16, P18) stated that they currently have a \textit{``go-bag''} or emergency kit prepared in case of emergencies. P08 described the details of her \textit{``go-bag''} and why she prepared such a bag:
\begin{quotation}
\textbf{P08}: \textit{``I have a go-bag... I put a month's worth of medication...And I update my medication...I rotate it to make sure I have a month's worth of medication...I have three days of clothing, because I figure you can wear it, you can rinse out, whatever. I have the little packets of Tide for washing. I have everything in this one bag...So many things that you don't think about, but I learned this after 9/11 when the government came out with, I think it was FEMA [who] recommended that everyone have a go-bag no matter where you live.''}
\end{quotation}

In describing these preparations, our participants emphasized the importance of knowing when emergencies can happen. While participants identified many tools they would like in a crisis app, almost all (17 out of the 18 participants), chose \textit{``emergency and weather alerts''} as their most preferred tool. Our participants further described how this feature must be \textit{``location sensitive''} and \textit{``up-to-date''} so that preparations can be made in advance and emergency assistance can be provided without delay. 

The importance of receiving alerts about upcoming crisis events were further explained by P02: \textit{``I had to know what's coming so that I can best determine how to prepare for it...as it impacts my action.''}. \hlfix{P02's description of wanting to take proactive action exemplifies participants' desires for crisis app features that support their sense of control in disaster situations. While the desire to know about potential hazards (\textit{``know what's coming''}) is reflective of a high levels of personal mastery reported by our participants (see \autoref{fig:ratings_control} and \autoref{sec:method_survey}), being informed about steps to stay safe helps them navigate through disaster, thus lowering their perceived constraints. In addition, since a heightened sense of control helps enhance preparedness among older adults during disasters, features that assist in maintaining this sense of control were preferred as these functions can further strengthen their awareness of emergency preparedness~\cite{timalsina2020factors}.}{Rewritten to address meta-review \#4 and R1's comment \#1} 

Nonetheless, the alert features chosen by participants may be purposeless if the users do not enable or unintentionally disable the location services for a crisis app. One participant described how this issue is common among older adults.
\begin{quotation}
\textbf{P06}: \textit{``What you're saying about location...A lot of [older] people don't realize it, that if you don't have your GPS turned on on your phone, you will not get any of these alerts. Because there's no way for them to know where you are and the location you're in. So when you go on your settings on your cell phone, you have to make sure that where it says GPS, that that's turned on. Otherwise, you're not going to get any service.''}
\end{quotation}

Another key preparedness strategy noted by our participants includes making sure that they are reachable by emergency managers. Beyond the toll that isolation can have on the mental health of older adults~\cite{brooke2020older}, isolation can be dangerous if emergency responders can not reach older adults quickly during emergencies. Some participants (P04, P07, P09) mentioned that their enrollment in emergency notification systems in their towns are part of their preparedness strategy and how the enrollment made them feel safe and secure. P04 described the utility of the enrollment:
\begin{quotation}
\textbf{P04}: \textit{``You can sign up to be on this emergency [notification system]...they do call us if there's some sort of emergency...I have a cellphone, and they called that...so they called when the power was out, when they said just to be aware if you need to go somewhere. I signed up for it when I was widowed...It's another peace of mind, really. Knowing that if something should happen...''} 
\end{quotation}

The actions described by our participants, whether preparing an emergency kit or enrolling in an emergency notification system, are reflective of the preparedness element of community resilience. \hlfix{While their preparedness actions may be induced by their knowledge and awareness of their situated vulnerability, these actions, focused on taking charge of their own safety and well-being, may also stem from participants' high sense of control. This is reflected through the fact that participants wanted as much information as possible about weather and emergency alerts from crisis apps to be well-informed (e.g., P02), indicating high personal mastery, and equipping themselves with emergency kits (P08) or generators (P06), indicating low perceived constraints.}{Added to address meta-review \#4 and R1's comment \#1} 

%\hlfix{In the next section, we present our findings about how participants' preferences for crisis app tools support active community participation, exemplifying the element of community relationships.}{Added to address R3's comment \#7 on signposting}

\subsection{Enhancing Active Community Role}

\hlfix{While prior work has identified multiple ways for older adults to participate in community-based activities and interpersonal interactions}{Rewritten to address meta-review \#6 and R3's comment \#9}~\cite{aroogh2020social}, it has not addressed how such participation can enhance older adults' community resilience.

\hlfix{Paralleling the high levels of social relationships reported in the survey, 14 out of the 18 participants chose the feature of being able to join a group with friends and family as important for the crisis app}{Added to address meta-review \#4 and R1's comment \#1} \hlfix{(see \autoref{fig:ratings_contact})}{Replaced Table 1 and address R1's minor comment \#2}. Participants also highlighted the importance of similarly being able to connect with neighbors and community members so that they can be aware of each others' well-being. In particular, P12 stated that \textit{``Well, if you are not at your home... and you can communicate with your neighborhood community, maybe somebody [who is] still at home, and can say, 'Your house looks okay.'''} P12's description of wanting to connect with neighbors during emergencies is an example of how older adults depend to those living close to them and provide support to others. 

Other participants described how being able to connect with community groups they participate in, such as \textit{``writing groups''} (P08) and \textit{``walking groups''} (P05) can be an extension to the support provided by a crisis app. They stated how such connections have made them feel a sense of belonging to the extent that other group members \textit{``checked up''} on them, especially when they are absent from group events. 

These relationships have increased the opportunities for older adults to play an even more active role in their communities during crises. For example, P11 described the help she provided to her neighbor during a tornado. 
\begin{quotation}
    \textbf{P11}: \textit{``My neighbor was up on a ladder repairing a window on the second floor when I was sitting in the house and got the [tornado] warning... It was raining at some point, and I said to my husband we have to get down in the basement, and then I remembered seeing my neighbor, so I went out the front door...and I started yelling to [my neighbor]...so about a month later, I saw [my neighbor] and [my neighbor] stopped and said...`You saved my life'.''} 
\end{quotation}

\hlfix{P11's description highlights the active role played by our participants in their community}{Rewritten to address meta-review \#6 and R3's comment \#9} simply by notifying others about upcoming emergencies, so together they can make preparations to stay safe. \hlfix{In parallel, these findings highlighted how older adults are critical resources for providing support and assistance to their community members, a contrast to the findings in previous work~\cite{tuohy2016older}.}{Rewritten to address meta-review \#6 and R3's comment \#9} By facilitating these opportunities, through technology or otherwise, older adults can continue to support their communities and be a beacon to provide others with assistance. In other words, a feature that facilitates informing neighbors of emergency alerts and sharing other pertinent information about their well-being or safety was identified as a key feature of crisis informatics tools that would enhance community resilience.

Beyond notifying others about emergencies, a group formed through the crisis app can be useful in identifying those who have the means and methods to shelter-in-place during emergencies. For example, P01 described how if she had such a crisis app, she would be able to know if someone close to her house had a generator during a power outage. P10 continued the conversation by explaining how a group would be useful for her.
\begin{quotation}
\textbf{P10}:\textit{``Well if your electricity is out it's nice to know who else's electricity is out. Because we have neighbors we'll call and say, ``Do you have electricity?'' Just so you know how far down... And I know a girl that lives over... And I know when hers is out that it's a bigger problem than if just my next door neighbor's out, because I know the line she's on. If only [the main area] is out, it's a big deal, because it's coming from the town. If we're both out, we know it's a transmitter somewhere and they're just going to fix that.''}
\end{quotation}

P10's description of the benefits of forming and joining groups in a crisis app demonstrates another way community resilience, through community relationships, can be developed among older adults and how such relationships can assist them in playing active roles in their communities. \hlfix{Correspondingly, the community relationships (with neighbors) described by P10 and P11 may also relate to their sense of control, instilled by becoming involved in community activities through the creation of a ``virtuous circle''~\cite{thornley2015building}.}{Added to address meta-review \#4 and R1's comment \#1} 

While we describe how these community relationships can occur between older adults, \hlfix{in the next section, we describe how community relationships are developed with emergency responders and how effective communication through crisis apps can support this relationship.}{Highlighted to address R3's comment \#7 on signposting}

\subsection{Accessing Emergency Support and Networks}

\hlfix{As shown in \autoref{fig:storyboard}, the storyboard prompted participants to list individuals or organizations that they felt comfortable receiving emergency assistance from. Contrary to previous studies which found that older adults would prefer getting help from friends and family~\cite{tuohy2016older, zhang2020understanding}, our participants indicated that they highly preferred support from local emergency services, despite their high social connectedness with friends and family (see \autoref{fig:ratings_contact}). Our participants discussed how their independent living conditions, living in their own homes rather than group settings, have persuaded them to be more reliant on government support services which are closer to them than family members and have relevant knowledge and expertise.}{Added to address meta-review \#4 and R1's comment \#1} Correspondingly, they focus on creating deeper connections with local emergency response agencies. For example, P03 described how the effectiveness of her local emergency responders (EMT) made her reliant on them:
\begin{quotation}
\textbf{P03}:\textit{``I'd end up calling the emergency person out. They are unbelievable... because the other people are worried about themselves. They never make you feel [like you are one of] a million people having problems... Because it doesn't even have to be the ambulance. They could call the community center and get a transport to you.''}
\end{quotation}

Similar to P03, P13 explained that he chose to have emergency responders as his primary support due to the convincing conversation he had with the local fire department. He was told to \textit{``anticipate''} the assistance he would need so that assistance can be provided to him without delay. 

Differing from P03 and P13, P01 and P17's confidence in their local police and fire departments stem from the constant communication that was provided during past emergencies. Specifically, P01 stated that local responders were proactive in contacting the community during severe events and therefore made her engaged with them. As for P17, she described her experience with the local police:
\begin{quotation}
\textbf{P17}: \textit{``The police are very good about calling if it's electricity, a storm, or a surf warning. They also are very good about telling you the time frame and keep you updated. So, they not only give you the information but they tell you what to expect and what to do in terms of, particularly, [when] we have a lot of wires that come down. Not to go near them. They give you what to do and the time frame on all of those. They do give you a call each time.''}  
\end{quotation}

\hlfix{While the reliability of emergency responders led our participants to prefer them over family and friends for receiving assistance during emergencies, participants' preferences may also reflect their sense of control. The emergency responders highlighted by P13 and P17, police and fire personnel, are resources that can be accessed and reached out to for fulfilling their emergency needs, which is linked to older adults' sense of control and independence~\cite{hillcoat2014meaning}.}{Added to address meta-review \#4 and R1's comment \#1} 

Other than support from services mentioned above, our participants described the importance of getting information about emergency shelters. Even though sheltering at home is preferred, our participants wanted information about shelters beforehand, especially when public buildings are utilized. For them, knowing the location of the shelters will enable them to make preparations as to the items they will need during a potential stay. P09 raised multiple questions to consider when learning about shelter locations: \textit{``Do they have beds? Do they have blankets? Do I need to bring anything? Do they have plugs you can charge your cell phone? Can I bring my own pillow? Will there be food provided?''}. She then stated that the answers to these questions \textit{``...would be helpful to know so you could plan accordingly.''}

Other than the location, knowing the current capacity of the shelters can also be helpful as the crowding, noise, and lack of privacy in a shelter may increase one's mental and physical burden~\cite{aldrich2008disaster}. Our participants discussed how crowding at shelters often creates more problems. One participant described her experience in the shelter:
\begin{quotation}
\textbf{P12}: \textit{``Sometimes, you go to a shelter though and, I don't know how many people have been in a shelter. But because we had to leave, we went to the shelter and we were in one room. But everybody was being brought in because the whole of, one half of [the] hall was underwater. The other wasn't. I'm going to tell you it's quite an experience and I don't know how many people would remain in a shelter if they were really put there. We left the shelter the next day. I signed out. I had to sign a paper that I was signing out. It was beyond chaos. But you took that responsibility upon yourself.''}
\end{quotation}

Other than location and capacity of emergency shelters, two participants (P06, P08) appreciated knowing if the shelters allow occupants to bring along their pets. The preferences mentioned by P06 and P08 are particularly important following a study in Miami-Dade County that found 35\% of older adults have pets and reported needing pet evacuation assistance~\cite{douglas2017pet}. The same study found that the low number of pet-friendly shelters in the county has increased older adults' vulnerability during Florida's hurricane season~\cite{douglas2017pet}. 

Based on the feedback from our participants, there are multiple opportunities for crisis apps to enhance support for older adults in their protective decision-making, whether in seeking emergency services or emergency shelters. As P07 said, \textit{``It helps you prepare for what you're doing while you have the opportunity to prepare for it.''}

\subsection{Trust and Agency in Crisis Apps}

We presented how the elements of community resilience in terms of local knowledge, preparedness, community relationships, and communication can be enhanced with older adults' preferences on crisis informatics tools. Some of their preferences were in part parallel to our earlier findings on their sense of control and social relationships. Here, we describe our findings on participants with knowledge and experience with technology (i.e.,``digitally ready'') were willing to engage with crisis apps following their trusts and agencies in the applications.

\subsubsection{Submitting to Cost of Personal Information}

To our knowledge, no prior work has explored the compromise older adults have to make in receiving support and services during emergencies. Specifically, enrollment in emergency management systems can include providing personal information that is often accessible through healthcare providers and protected by Health Insurance Portability and Accountability Act or HIPAA~\cite{hipaa2021}. For example, New Hanover County in North Carolina requires a long list of information from their \textit{special needs} constituents when signing up for their emergency management system~\cite{newhanover2021}. For some users, providing this information make them vulnerable to possible fraud and may threaten their privacy. But, without the critical and timely crisis and risk information, support during emergencies may be delayed, especially when medical attention needs to be provided. We explore the critical need of information for emergencies by seeking to understand which types of information our participants were willing to provide on a crisis app.

Beyond general information such as name, phone number, emergency contact, and address, all participants were in agreement that they were comfortable providing details about their primary care physicians and pharmacies. Some participants described the traditional way of posting the information on their refrigerator but acknowledged the importance of a trusted third-party, such as the emergency managers, having the information as well. One participant explained why providing the information would help during emergencies:
\begin{quotation}
\textbf{P09}: \textit{``I think the primary care physician is a good thing because I think a place that would have your patient number included on that. And then they could just say, `We're with patient number...'  Whatever your number is. And then they can go into your history... Hopefully someone goes in and can say her blood type...You know they have everything. And that's the reason why this kind of information is important as well is when evacuation really happens and you don't have enough medicines for yourself, you want to make sure that you have those medications...or maybe the closest center has you medication on hand so they can actually give it to you. Because one of the reasons are medication.''}
\end{quotation}

Participants also discussed providing details of home health care and medical equipment providers for those who utilize them. They also noted that other older adults they knew who had caregivers at home, and that providing this information may help emergency response processes. Interestingly, only one participant supported providing information that is more specific to a person, such as eye color, height, and weight. Other participants described \textit{``not seeing the reason''} to provide these information, but one participant, P08, stated how the information is helpful in making sure the providers are not assisting the wrong person. 

Our findings in identifying the kinds of information that older adults are willing to provide can help emergency managers develop crisis apps that support both the entry and privacy of the information. Ultimately, the usefulness of a crisis app is contingent upon the use by the individuals who provide information and the managers who utilize the information. Some information, such as eye color, may seem required by responders but may not be seen as mandatory from the perspectives of the users. As such, it is important to examine to what extent the perceptions of the usefulness of ``required'' information in the crisis apps are aligned between information providers (i.e., older adults) and information requesters. The discrepancies in  the perceptions of different stakeholders should provide important implications for design. Therefore, underlining information that is required and optional may encourage older adults to enroll in the crisis apps while supporting the planning for emergency response deployment. In our study, participants highlighted the benefits of providing the information, as described by P18: \textit{``You're making their jobs harder if you don't give them the tools that they need to help you.''}

\subsubsection{Subscribing to Crisis Apps}

One common theme that featured in crisis informatics research is the reliability of crisis apps~\cite{zhang2020understanding}. Our study did similarly addressed this theme. Throughout the group discussions, our participants discussed their trust in mobile applications, especially since our storyboard prompted their willingness to submit personal information through a crisis app. In particular, P03 described how her willingness to provide information must not be confused with the willingness to subscribe to any kind of crisis apps. She wanted to know who and where her information was going and the security of the system that handles her information. The requirements were further explained by another participant:
\begin{quotation}
\textbf{P06}:\textit{``I think who also sponsors the app. Who is the seller of the app? I know I would be leery about downloading something until I know, number one, does it work, but also number two, the reputation. What is this group? Who is this group? Yeah, and what are they going to do with that information? ...I'm equating it to something very similar to Facebook. Very innocently [until] it's exploded in what it's become, but also the vulnerability that people have, and without even realizing it.''}
\end{quotation}

On the other hand, some participants (P05, P07, P08) described how crisis apps should be maintained by a single trusted entity since the information from multiple sources can be confusing to them when trying to decide the best protective actions to take. In particular, P07 described his experience with conflicting information from multiple sources during 9/11 were a result of misinformation, which creates more confusion than assistance. In describing the trusted entity, P05 described that support for the app by a trusted organization, such as senior centers, could improve adoption:
\begin{quotation}
\textbf{P05}: \textit{``See, I would like to think that the senior center here, if that [crisis app] became available...simply say, "After some research, we feel comfortable that this could be a potential tool that could be useful in the future." That would give me a little bit of warm and fuzzy feeling.''}
\end{quotation}

The above findings highlight participants' willingness to engage in crisis apps by providing personal information for the sake of safety. The findings that relate to privacy concerns were one of the most common barriers found in crisis apps intervention, especially among older adults~\cite{kadylak2020united}. Participants described how the barriers could be overcome by identifying trusted agencies that deploy the crisis apps.

%------------------------------------------------------------

\section{Discussion}
Our findings illuminated how crisis apps can be utilized to build and enhance community resilience during emergencies. This is described through participants' choices of tools and features that they deemed important to help them during emergencies. While chosen based on the needs of participants, these features are aligned with the elements of community resilience~\cite{patel2017we}. We conclude with a discussion encapsulating the interplay of community resilience, crisis informatics tools, and older adults. Further, we discuss opportunities for supporting older adults' usage of crisis apps.  

\subsection{Enhancing Community Resilience via Crisis Informatics Tools for Older Adults}

Prior work has found that the core to community resilience is local knowledge, preparedness, community relationships, and communication~\cite{patel2017we}. Our findings suggest how these elements are enhanced with older adults' preferences for crisis apps. For example, we describe local knowledge pertaining to a disaster context through situated vulnerability, or vulnerability related to the residing location. Our findings show that awareness and knowledge about the risks and hazards associated with the geographical location influence how older adults prepare for crisis events. Our participants discussed how crisis apps that notify about severe weather and emergency forecasts, would enable them to take protective actions and make decisions pertaining to their safety. While tools that notify about upcoming disasters may be common in crisis apps~\cite{tan2017mobile, zhang2020understanding}, an understanding of how these tools can enhance community resilience must be continually addressed. For example, these same tools may be accompanied by messages from local emergency responders providing older adults with tips and guidance for preparations, ultimately reducing negative outcomes from lack of preparedness~\cite{tuohy2016older, shih2018improving}. 

Preparedness, as described in our study, also helped participants build community resilience as it included utilizing community resources (e.g., emergency notification systems). While studies have mentioned the importance of preparedness in building community resilience, few described the types of preparations involved in mitigating the effects of disasters~\cite{patel2017we}. Our findings described the details of preparedness actions taken by older adults residing in coastal locations. These preparedness strategies, while easy to deploy (e.g., preparing an emergency kit), remain unexploited by many older adults~\cite{zhang2020understanding}. \hlfix{Through crisis apps, customized tips can be made available for older adults so that they can better prepare for their unique conditions (e.g., climate, altitude, etc.)}{Revised to remove negative stereotypes and address meta-review \#6 and R3's comment \#9}. For example, older adults residing in areas prone to wildfires may require respirators in their emergency kits~\cite{ready2021FEMA}. The preparation for planning and mitigation measures can support sustainable response and recovery by the community and reduce the likelihood of harm and danger to older adult communities. 

\hlfix{Similarly, both local knowledge and preparedness may be related to concepts of the Protection-Motivation Theory~\cite{rogers1975protection}, which helps explain how individuals are motivated to react in a protective manner in response to a perceived threat~\cite{westcott2017expanding}. Previous studies that have utilized this theory in disasters contexts, such as floods~\cite{botzen2019adoption}, bushfires~\cite{westcott2017expanding}, and the COVID-19 pandemic~\cite{kim2020age}, have focused on the general population. Therefore, future work may consider using Protection-Motivation Theory to examine the role of technology in enhancing community resilience, and to explore how technology can support marginalized communities (e.g., older adults) in their threat and coping appraisal.}{Added to address meta-review \#7}

The importance of community relationships between older adults and emergency responders was highlighted in the emergency support preferences of the participants in our study. Other studies report that older adults often seek support through familial connections during emergencies~\cite{ashida2017personal,cohen2016community}. For example, Ashida et al. stated that the ability to interact frequently with family and friends is an important implication for the availability and accessibility of support sources during emergencies~\cite{ashida2017personal}. Their study further explained that the frequency of interactions via phone or internet was shown to be an important predictor of being selected as a source of emergency support. \hlfix{However, in our study, we found that older adults with a high frequency of social contact--with family and friends via all kinds of interactions (i.e., face-to-face, video, phone, text messages, etc.)-- still prefer to receive support from professional emergency responders who have knowledge and expertise to assist them during crises. Even though expecting that they need help may suggest high perceived constraints (contrary to \autoref{fig:ratings_control}), actively choosing who they want to receive assistance from represents a high personal mastery.}{Highlighted to address meta-review \#4 and R1's comment \#1} When coupled with a computational artifact that provides a critical support function for older adults, these relationships enhance the community resilience factors that are essential in disaster response efforts~\cite{leykin2013conjoint}. \hlfix{For example, our participants described how crisis apps that featured the ability to form and join specific groups could support their decisions and actions during emergencies.}{Revised to address older adults in active term rather than being helped, meta-review \#6 and R3's comment \#9} They sought to join groups with those living nearby and who they have established community relationships with.

The efforts for facilitating community relationships must then be coupled with communication protocols that enable older adults to not only support themselves but each other. Our findings show how older adults play an active role in the community by sharing critical information and providing support and assistance, in ways as simple as sharing at-home generators. Previous studies have shown how power outages have led to worsening conditions for older adults and how these findings do not significantly differ between older adults who live alone or with others~\cite{bell2020predictors}. Through the ability to communicate with personalized groups, older adults are empowered to engage with their communities, reducing social isolation and loneliness~\cite{gray2018supporting}. More importantly, by supporting each other, older adults can become a beacon for emergency managers to contact isolated members of the community. For example, they can influence others to enroll in emergency notification systems or to engage with community organizations. When deeper connections are made, the community grows together to become more resilient, especially during crisis events. 

While we address the needs and preferences of older adults that parallel with the elements of community resilience, we acknowledge how community resilience itself includes the participation of organizational leadership. Elements of local knowledge, preparedness, community relationships, and communication need to be supported by the organizations that deploy the crisis apps since community resilience itself is constrained by both the abilities of individuals and collective efforts to achieve desired ends \cite{rand2011building}. Nonetheless, by addressing how community resilience can be enhanced through the design and deployment of crisis apps from the perspectives of older adults, our findings call for attention to structural leadership conditions, given the importance of resource efficacy as described by our participants. Therefore, future work should include engagement with emergency managers, particularly those in coastal locations, to identify their attitudes and preferences for crisis apps that seek to build and enhance community resilience with older adults.

\subsection{Opportunities and Challenges for Crisis Informatics}

Through the examination of older adults' perceptions about the utility of crisis informatics tools in enhancing community resilience, we identified several opportunities and challenges related to the development of these tools. First, we highlighted opportunities for crisis informatics tools to include information regarding shelter locations, features, and availability. Past studies about disaster preparedness and implications for older adults have discussed how older adults avoid staying in shelters during emergencies due to overcrowding, noise, and lack of privacy~\cite{aldrich2008disaster}. Our findings show that providing information about emergency shelters in crisis apps, particularly when coupled with connections to their personalized groups, can provide the opportunity for older adults to exercise their sense of autonomy by supporting older adults' decision-making and actions for protecting themselves.

Our identification of preferred and required services addresses a critical gap in the literature, specifically on the lack of understanding about the support and networks required by older adults during emergencies. Beyond identifying the important features, we additionally characterized the willingness of participants to share information to facilitate receiving their preferred support services. Having this information during the emergency preparedness phase helps emergency managers delegate rescue and support tasks more efficiently. In fact, decisions without reference to vital information are often made from failure to have pertinent information in the first place~\cite{sharps2002mindless}. The utility in gathering information to make early decisions for saving lives will significantly help an unprecedented worsening situation.  Ultimately, it is the joint inclusions of both the most effective features and the information sharing by participants that will make crisis apps most beneficial for emergency responders and older adults.

Nonetheless, collecting users' information for the purposes of emergency support requires the trust in both the information seeker and the crisis platforms. In our case, the seeker is the emergency responders, and the trust can be gained through regular contact and connections as described by our participants. Our findings identify how participants have a high level of trust with their fire and police departments due to the departments' reliability in delivering information, help, and support. While this process may differ from one location to another, our findings address the key needs to create an appropriate communication infrastructure and to implement strategies that can be coordinated in disaster settings. Further, we identified the importance of users' having trust in the privacy and security features of the crisis apps to facilitate broader adoption of this technology. Based on these findings, we encourage future work to explore additional strategies to gain older adults' trust in crisis apps. Without a sense of trust, well-developed crisis apps may be under-utilized, limiting the promise of crisis apps for older adults. 

\hlfix{Finally, our findings demonstrate the need for crisis informatics to not only support older adults to prepare for disasters, but also to connect them with their community as a whole, which include their nearby family and friends, neighbors, and emergency responders. The role of crisis informatics tools in achieving community resilience and supporting older adults during disasters are immense, not only by facilitating these connections, but also by providing support beyond information, such as tangible, emotional, and esteem support. For example, our participants described how community relationships built through crisis apps are useful for sharing access to critical resources (e.g., generators), providing assistance (e.g., checking on status of their house), or comforting each other (e.g., checking on well-being). Specifically, participating in community groups in web-based spaces (i.e., crisis apps), can foster offline interactions between older adults~\cite{nurain2021hugging, vzivst2021deployment}. Extensions of this work can include an increased focus on the use of crisis apps in addressing these other types of support. Further, since the strength of older adults comes partially from their accrued experiences with multiple types of crisis events~\cite{shih2018improving,lind2021more,yuan2019beyond, soro2015noticeboard}, another extension of this work should examine and compare these findings among age groups.}{Added to address meta-review \#5, R1's comments \#2, R3's comment \#8}

\subsection{Limitations and Future Work}

Our work involved older adults with a heightened sense of control, established social relationships, and digital readiness. While we underscore the needs and preferences of these types of older adults, we acknowledge how these identifications may not necessarily reflect the diversity of needs within the older adult population. For example, our participants did not require assistance with activities of daily living (ADL). By contrast, older adults with severely limiting physical conditions may not have the same level of sense of control~\cite{barusch2011disaster}, familiarity with technology~\cite{arevian2018participatory} and community engagement~\cite{ward2012sense}. Therefore, our findings must be interpreted cautiously. It is also possible that some participants over-reported or under-reported their sense of control, social relationships, and digital readiness. Nonetheless, the projection of growth in the older adult population in the future~\cite{world2019united} is likely to include growth in cohorts with high levels of technological knowledge~\cite{pruchno2019technology} and potential readiness to engage in their communities. Thus, we must be forward-thinking in understanding how these shifts will change methods of assisting older adults in the future, particularly during crises.

\hlfix{Moreover, we reflected on how the timing of our study might have influenced our findings and design implications. Recall that our study took place before COVID-19 was declared a national emergency in the United States~\cite{whitehouse2020covid}. Thus, our findings are limited to the understanding of opportunities for technologies that promote community resilience in a world without social distancing requirements. In the aftermath of the implementation of COVID-19 restrictions, researchers have sought to identify strategies to promote community connections among older adults~\cite{van2021strategies} while social distancing. Thus, crisis informatics research may consider investigating the implications of these strategies for future systems, especially when limits on in-person meetings with their community networks may occur again. Such research may include understanding how critical information about significant events is exchanged between older adults to keep each other informed, while continuously ensuring their well-being.}{Added to address meta-review \#3}

\balance 

Additionally, we chose to engage older adults with digital readiness as key informants that could help us investigate opportunities and challenges that this design space represents. We specifically sought this attribute when advertising our study as we project that future older adults will have similar attributes, given the evolving engagement of this population with technology~\cite{pruchno2019technology,pew2021internet}. Still, we note that soliciting the perspectives of older adults with low digital readiness is an equally important endeavor for future work to ensure that a diverse range of perspectives are used to drive the design of future systems. 

\hlfix{Indeed, amidst the COVID-19 pandemic, during which many older adults expanded their technology use to combat loneliness during the nationwide lockdown in the United States~\cite{bloomberg2020} and around the world~\cite{sin2021digital,von2020smart}, we have observed increased digital readiness among older adults. However, reports have also documented the technological difficulties experienced by this population during the pandemic. For example, researchers have described the challenges that older adults have experienced when attempting to register online for COVID-19 vaccine appointments~\cite{nyt2021older,theverge2021older}. These contradictory findings highlight that while many adults are increasingly engaging with ICTs, technological systems are not yet meeting their needs amidst a crisis.}{Rewritten to address meta-review \#3} \hlfix{However, while we recruited participants who use digital technology on a daily basis, we did not assess participants' level of experience with existing crisis apps. Future work may consider focusing on older adults who have utilized crisis apps to examine their experiences with such platforms, and how these experiences shape their desires for future tools. We refer readers to the work by Zhang et al.~\cite{zhang2020understanding} on how existing crisis apps may support older adults' sense of control, safety, and dignity.}{Added to address R1's comment \#4}

In the future, we aim to include a more diverse older adult population spanning different races, levels of education, and income levels, to understand how these factors influence their disaster strategies, and how community resilience among older adults from different backgrounds can be established and supported through crisis informatics tools. While none of the authors are in the ``older adult'' cohort, in the future, to further ensure that older adults' perspectives are appropriately and fully reflected in the analysis process, it may be beneficial to include older adults as members of the research team or as part of a community advisory board, and to engage in member checking of the data with older adults.

%------------------------------------------------------------

\section{Conclusion}

The increasingly common natural disasters, public health emergencies, and other crisis scenarios inform a serious need to examine how we can better empower our older adult population. Our findings shed light on opportunities for crisis informatics tools to promote community resilience not only among older adults, but also with their community members and emergency personnel through the elements of local knowledge, preparedness, community relationships, and communication.

\begin{acks}
We thank our participants for sharing their stories and experiences with us and the recruitment site for assisting us in the recruitment process. We also thank our anonymous reviewers for their insightful comments to strengthen the paper. This study was supported by Northeastern University TIER 1 Seed Grant/Proof of Concept Program.
\end{acks}

%% The next two lines define the bibliography style to be used, and the bibliography file.
\bibliographystyle{ACM-Reference-Format}
\bibliography{ref}

\end{document}